\documentclass{aa}  

\usepackage{graphicx}
\usepackage{txfonts}
\usepackage{hyperref}

\usepackage{todonotes}
\usepackage{xspace}

\newcommand{\angstrom}{\text{\normalfont\AA}}

\newcommand{\HI}{{\text{H\MakeUppercase{\romannumeral 1}}}\xspace}

\newcommand{\Lya}{\ifmmode{\mathrm{Ly}\alpha}\else Ly$\alpha$\xspace\fi}
\newcommand{\kms}{\,\ifmmode{\mathrm{km}\,\mathrm{s}^{-1}}\else km\,s${}^{-1}$\fi\xspace}

\newcommand{\npeaks}{\ifmmode{N_{\mathrm{peaks}}}\else $N_{\mathrm{peaks}}$\xspace\fi}

\makeatletter
\renewcommand*\aa@pageof{, page \thepage{} of \pageref*{LastPage}}
\makeatother

\begin{document}

   \title{Variations of Observed Lyman-$\alpha$ Spectra Shapes due to the Intergalactic Absorption}

   \author{C. Byrohl
          \inst{1}
          \and
          M. Gronke\inst{2,3}\thanks{Hubble fellow}
          }
          
   \institute{Max-Planck-Institut für Astrophysik, 
              Karl-Schwarzschild-Str. 1, 85748 Garching, Germany \\
              \email{cbyrohl@mpa-garching.mpg.de}
         \and
         Department of Physics and Astronomy, Johns Hopkins University, Baltimore, MD 21218, USA
         \and
         Department of Physics and Astronomy, University of California, Santa Barbara, CA 93106, USA
             }

   \date{Received MONTH DAY, YEAR; accepted MONTH DAY, YEAR}

\abstract
{
    Lyman-$\alpha$ (Ly$\alpha$) spectra provide insights into the small-scale structure and kinematics of neutral hydrogen (HI) within galaxies as well as the ionization state of the intergalactic medium (IGM). The former defines the intrinsic spectrum of a galaxy, which is modified by the latter. These two effects are degenerate.
    Using the IllustrisTNG100 simulation, we study the impact of the IGM on Ly$\alpha$ spectral shapes between $z\sim 0$ and $5$. We compute the distribution of the expected Ly$\alpha$ peaks and of the peak asymmetry for different intrinsic spectra, redshifts and large-scale environments.
    We find that the commonly used \textit{averaged} transmission curves give an incorrect perception of the observed spectral properties. We show that the distributions of peak counts and asymmetry can lift the degeneracy between the intrinsic spectrum and the IGM absorption. 
    For example, we expect a significant number of triple peaked Ly$\alpha$ spectra (up to 30\% at $z\sim 3$) if the galaxies' HI distribution become more porous at higher redshift as predicted by cosmological simulations. 
    We provide a public catalog of transmission curves to be used in future simulations and observations to allow a more realistic IGM treatment.
  }

   \keywords{Radiative transfer --
                intergalactic medium --
                large-scale structure of Universe -- Galaxies: high-redshift
               }

   \maketitle

\section{Introduction}
\label{sec:intro}

The Lyman-$\alpha$ (\Lya) line is a promising astrophysical observable for the neutral
hydrogen distribution, from the scale of parsecs in star-forming regions
\citep[e.g.,][]{1998A&A...334...11K,2016ApJ...820..130Y} all the way to cosmological scales. As such, it is among the most powerful observables to constrain the cosmic neutral fraction during the ``Epoch of Reionization''
\citep[e.g.,][]{2014PASA...31...40D,2019MNRAS.485.3947M}.

While \Lya observations allow us to tackle a wide range
of astrophysical questions, probing different spatial scales is a challenge because of a possible degeneracy of the origin of spectral features in those observations.
\Lya observables are shaped not only through the neutral hydrogen distribution
and kinematics internal to galaxies (i.e., the interstellar medium, short ISM) but also by neutral
hydrogen residing in the circumgalactic medium (CGM) \citep{2011ApJ...736..160S,2016A&A...587A..98W} and the intergalactic medium (IGM).

At low redshift ($z\lesssim 1$), the IGM's impact is very limited due to low neutral hydrogen fractions, such that observations with the Hubble Space Telescope reveal the spectral features imprinted by ISM and CGM only. Usually \Lya spectra show very little flux at line-center ($\lambda_{c}\sim 1216\ $\AA), and emission mostly originates on the red ($\lambda > \lambda_{c}$) side of the spectrum with
a significant fraction of spectra
showing essentially no flux on the blue ($\lambda < \lambda_{c}$) side~(\citealp{2014ApJ...797...11O,2014ApJ...782....6H,2015ApJ...809...19H,2016ApJ...820..130Y}; review by \citealp{2015PASA...32...27H}).

At higher redshifts, however, the impact of the IGM increases, and the picture is less certain. Individual observations at $z\sim 5-7$ show mostly a single peak redshifted by a few hundred $\kms$ \citep{2017MNRAS.472..772M}. At intermediate redshifts, larger statistical samples do measure an asymmetry towards the red \citep{Erb2014}. This spectral evolution leaves, in principle, two possibilities: either the `intrinsic' \Lya spectra emergent from the galaxies do not vary strongly with redshift, or the intrinsic \Lya spectra vary but the also evolving IGM transmission makes the observed spectral properties stay similar. While the former option is supported by the fact that the low-$z$ samples are selected to be `analogs' of higher redshift \Lya emitters \citep[e.g.,][]{2016ApJ...820..130Y}, the latter is suggested by modern radiative transfer simulations using galactic hydrodynamical simulations as input \citep[e.g.,][]{Laursen2011,2018ApJ...862L...7G,Smith2019}. Due to strong feedback mechanisms, they produce a porous ISM at high-$z$, and thus, the predicted \Lya spectra exhibit relatively large flux at line-center and the blue side of the spectrum. Differentiating between these pathways is crucial to properly disentangle the \Lya line's use as a probe of galaxy and IGM evolution.

While recent studies focused mainly on the intragalactic \Lya transfer \citep[e.g.,][]{Smith2019}, less attention has been attributed to the effect of the IGM on the \Lya spectral shape -- even large scale studies including the IGM focused primarily on global statistics such as the \Lya emitter clustering or luminosity function \citep[e.g.,][]{Iliev2008,Zheng2010,Behrens2018,Byrohl2019}.

In this paper, we seek to clarify the IGM's impact on the \Lya spectra using using a recent cosmological simulation in a redshift range $z=0-5$. 

\section{Methodology}

\subsection{Simulations}

We analyze the IGM attenuation using the IllustrisTNG100 simulations \citep{2018MNRAS.477.1206N,2018MNRAS.475..624N,2018MNRAS.480.5113M,2018MNRAS.475..648P,2018MNRAS.475..676S} with a box size of $106.5\ $comoving Mpc for redshifts $0.0$, $1.0$, $2.0$, $3.0$, $4.0$ and $5.0$.
The attenuation of \Lya flux is calculated with a modified version of \textit{ILTIS}\footnote{An earlier version of ILTIS is publicly available here: \url{https://github.com/cbehren/Iltis}}, a line emission transfer code as presented in \cite{Behrens2019}, tracing the optical depth in the IGM between the \Lya emitting galaxies and the observer for chosen lines of sight. The code has been modified to natively run on IllustrisTNG's Voronoi tessellation, eliminating the prior need for an intermediate interpolation step onto a suitable grid-based dataset structure. The initial Voronoi tessellation is created with a parallelized wrapper to the tessellation code \textit{voro++} \citep{voroplusplus} on IllustrisTNG's particle distribution. 
IllustrisTNG uses a time variable UV background with self-shielding \citep{2009ApJ...703.1416F,2013MNRAS.430.2427R} responsible for the hydrogen's ionization state in the IGM.

The optical depth $\tau$ is integrated over the intervening medium for given lines of sight and a given input wavelength $\lambda_{\mathrm{i}}$. The optical depth along the way can be expressed by the integral

\begin{align}
\label{eq:1}
    \tau(\lambda_{\mathrm{i}}) = \int_{s_0}^\infty \mathrm{d}s \ n_{\HI}(s)\ \sigma\left(\lambda(\lambda_{\mathrm{i}},v,s),T_{\rm HI} \right),
\end{align}
where $n_{\HI}$ is the neutral hydrogen density and $\sigma(\lambda)$ is the
corresponding cross-section for a photon interacting with a neutral hydrogen
atom. In equation~\eqref{eq:1} we integrate over the physical distance $s$ from the source along the chosen line of sight. The temperature $T_{\rm HI}$ sets the thermal broadening reshaping the cross-section profile $\sigma$. The cross-section is evaluated in the gas' rest-frame and thus depends on the peculiar velocity $v$ and Hubble flow $H(z)$ at redshift $z$. The wavelength is shifted as
$\lambda = \lambda_{\mathrm{i}}\left[1+\frac{v(s)+H(z)\cdot s}{c}\right]$, where $c$ is the speed of light. We commonly express the wavelength as its offset $\Delta\lambda_e=\lambda-\lambda_c$ from the \Lya line-center at the emitters' redshift.

The input wavelengths $\lambda_{\mathrm{i}}$ are evaluated in the rest-frame of the halos, which we identify as the mass-weighted velocity of the respective halo. We compute the optical depth within the wavelength range $\left[\lambda_c - 5\ \angstrom,\lambda_c + 3\ \angstrom\right]$ with a resolution of $0.02\ \angstrom$ ($R\sim 60000$). As we are interested in the large-scale attenuation, we start summing contributions to the optical depth from a distance $s_0=f r_\mathrm{vir}$, where $f$ is a factor and $r_\mathrm{vir}$ the virial radius of the halo. For comparison with \citet{Laursen2011}, we choose $f=1.5$. We integrate all attenuation contributions for distances up to $30\ $cMpc/h using periodic boundary conditions for the box. This length corresponds to a Hubble shift of 2600 km/s (at $z=1.0$) or more. We verified that for the chosen input wavelength range $\lambda_{\mathrm{i}}$, all wavelengths have significantly redshifted beyond the Ly$\alpha$ line-center as facilitated by the Hubble flow and thus no attenuation contributions are expected beyond this upper integration limit.

In our analysis, we consider the centers of halos as possible \Lya emitting galaxies if they contain regions of active star formation and have a total halo mass above $5\cdot 10^{9}\ $M$_\odot$ as provided by IllustrisTNG's Friends-of-Friends halo catalogs.

For each emitter, we evaluate the optical depth for the same set of 1000 lines of sight (LoS). The LoS are constructed as normal vectors of a 1000-faced Fibonacci sphere evenly tracing possible directions.

A reduced public data set of our transmission curves has been published as \citet{Byrohl2020dataset} and a full data set will be made available upon request.

\subsection{Input Spectra}
\label{sec:inputspectra}

To demonstrate how the IGM attenuation affects the observed spectra in a statistical sample using the individual attenuation curves, we need to assume some input (intrinsic) spectra with flux density $I_{\lambda,\mathrm{input}}$. Here, we use three different toy models:
\begin{enumerate}
    \item A symmetric double peaked profile as result of a static neutral
      hydrogen sphere (``Neufeld solution'') \citep{Neufeld1990,Dijkstra2006}. Here we set the temperature as $T_{\rm HI}=10^4\ $K and the column density to $N_{\rm HI}=10^{20}\ $cm$^{-2}$.
    \item A red peak only that corresponds to the Neufeld solution (see above)
      under the assumption of a significant outflow. Together with intrinsic profile \textbf{(1)} they bracket observed cases at low-$z$ which largely consist of a single or double peak dominant toward the red.
    \item A Gaussian at Ly$\alpha$ line-center with a width of $\sigma=200\
      $km/s. Such setup with a significant line-center flux can be motivated for
      galaxies with a larger impact of stellar feedback at high redshifts leaving a more
      porous HI distribution. \Lya photons then escape closer to
      line-center and are less susceptible to the gas kinematics
      \citep{Neufeld1991,HansenOh2006,2016ApJ...826...14G}. For this reason, recent galactic hydrodynamical models post-processed with \Lya radiative transfer show a wide, fairly symmetric intrinsic profile 
      with little absorption at line-center
      emergent from these galaxies \citep[e.g.,][]{Smith2019}. The width of the Gaussian was chosen to approximately match the predictions of those models.
\end{enumerate}

\section{Results}
\label{sec:results}

\subsection{Averaged Transmission Curves}
\label{sec:transmission}

\begin{figure}[hbt]
\centering
  \includegraphics[width=1.0\linewidth]{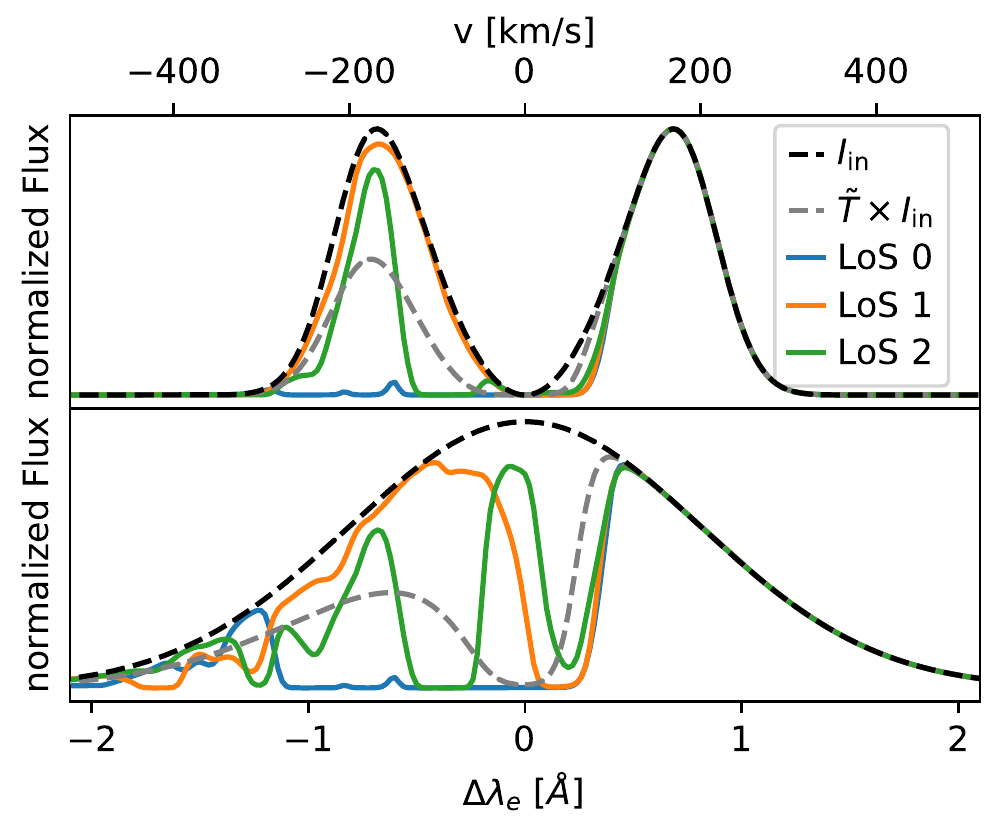}
  \caption{The panels show two different input (i.e. intrinsic) spectra (black lines) and how the IGM attenuation shapes the observed spectra along different LoS at $z=3.0$. The gray dashed line shows the multiplication of the respective input spectrum with the median transmission curve $\tilde{T}$ for all emitters and LoS at the given redshift. The colored solid lines show observed spectra for the same emitter but for different LoS. Wavelengths are evaluated in the halo's rest-frame. \textbf{(top)} The input spectrum is obtained as analytic solution of a spherical hydrogen distribution with a column density $N=10^{20}\ $cm$^{-2}$ and temperature $T=10^4\ $K \citep{Dijkstra2006}. \textbf{(bottom)} Using a Gaussian intrinsic spectrum with standard deviation $\sigma=200\ $km/s.}
\label{fig:individualSpectraLoS}
\end{figure}

In Figure~\ref{fig:individualSpectraLoS}, we show the resulting spectrum along three LoS for the same origin with the Gaussian and double-peaked input spectra. In this plot and in general, mostly the spectrum bluewards of the line-center is affected as those frequencies will eventually shift into the line-center by the Hubble flow. Figure~\ref{fig:individualSpectraLoS} shows that the transmission bluewards of the line-center can fluctuate strongly for different wavelengths of a given line of sight. In fact, the LoS in Figure~\ref{fig:individualSpectraLoS} have been chosen such that the blue side of the observed spectrum exhibits a varying count of spectral peaks between zero and two for the Neufeld input spectrum. We will formalize the count of peaks into a quantitative measure in \S~\ref{sec:Npeaks}.

The transmission function is given as
\begin{align}
    T(\Delta\lambda_e)=\exp\left[-\tau(\Delta\lambda_e)\right]
\end{align}
and describes the fraction of the overall flux attenuated by the IGM between the emitter and the observer.

\begin{figure}
\centering
  \includegraphics[width=1.0\linewidth]{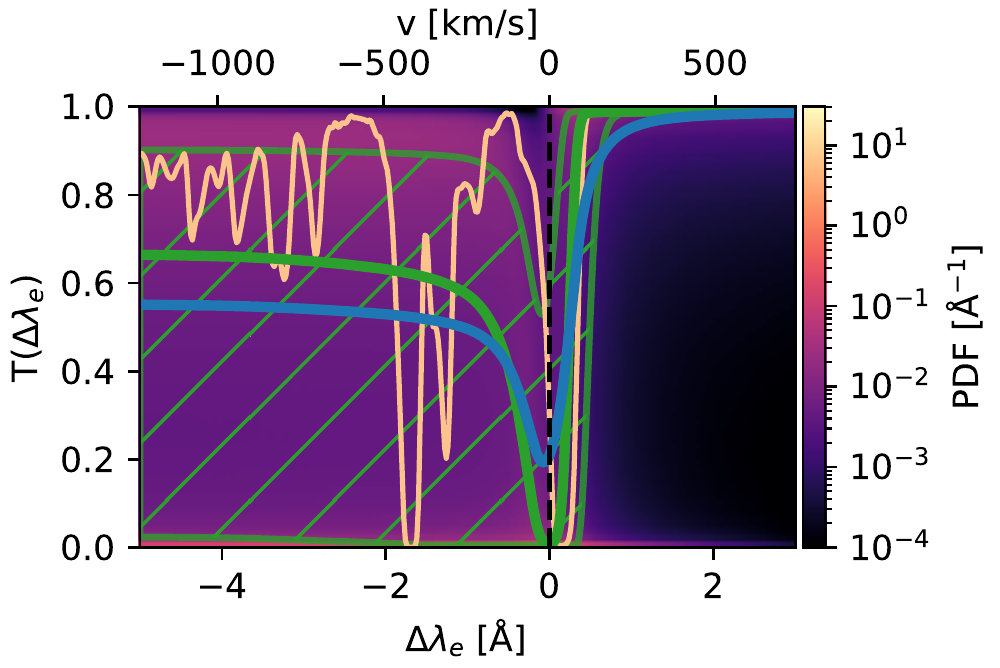}
  \caption{The PDF of the transmission $T$ as a function input wavelength shift $\Delta\lambda_e$ over all emitters and LoS at $z=3$. The green line shows the median transmission with the hatched region enclosing the central $68\%$ of all individual transmission curves, while the blue line shows the mean transmission curve over this PDF. The light orange curve shows an example of an individual LoS (corresponding to "LoS 1" in Figure~\ref{fig:individualSpectraLoS}).}
\label{fig:transmissionPDFz3}
\end{figure}

In Figure~\ref{fig:transmissionPDFz3}, we show the probability density function (PDF) $p(T,\Delta\lambda_e)$ of the transmission function $T$ averaged over all emitters and LoS at $z=3$. We also plot the mean and the median curves (blue/green bold lines) along with the central 68 percentiles (hatched area). From these curves, it is clear that the blue side is suppressed by roughly a factor of two, while the red side is mostly unaffected. We find a trough around line-center suppressing most of the flux. In general, we find these curves to be consistent with the results found by ~\citet{Laursen2011} and \citet{Lopez2020}. Discrepancies in the asymptotic median value and the shaded 16th-84th percentile region are mostly a result of these quantities being highly dependent on the spectral resolution (see Appendix~\ref{sec:spectralres}). The spectral resolution in the literature is significantly lower with $R\lesssim 3000$ than $R\sim 60000$ used here.

For other redshifts, we find a similar qualitative trend over the shown wavelength range: the transmission on the blue side is increasingly suppressed at higher redshifts with a smaller asymptotic value towards $\Delta\lambda_\mathrm{e}=-5\ $\AA\ and a deeper trough around  $\Delta\lambda_\mathrm{e}=0\ $\AA. We find the most likely offset for this central trough to be roughly $20\ $km/s at $z=0.0$, monotonically increasing towards $70\ $km/s at $z=5.0$. These velocity offsets and their redshift evolution are consistent with expected halo infall velocities at $r=1.5 r_{\mathrm{vir}}$~\citep{Barkana2004}. 

As illustrated by Figures~\ref{fig:individualSpectraLoS} \& \ref{fig:transmissionPDFz3}, in general, the median curve is misleading and should be interpreted with caution: The underlying PDF $p(T|\Delta\lambda_e)$ of transmission $T(\Delta\lambda_e)$ is usually bimodal, that is, due to the large \Lya cross section, mostly close to zero or unity -- which is ill represented by averaged transmission curves.
For instance, in Figure~\ref{fig:transmissionPDFz3} the bimodal distribution peaks around $0.0$ and $0.9$ on the blue side at $z=3.0$. While this bimodality becomes more pronounced at lower redshifts, a unimodal distribution with $\langle T(\Delta\lambda_{\rm e})\rangle\sim 0$ forms at higher redshifts as the upper bimodal transmission peak value decreases. For wavelengths slightly redwards of the line-center we can also see this bimodality strongly pronounced for the two transmission values zero and one.

This bimodal behaviour has important consequences for the observed spectra. Rather than blue peaks being uniformly suppressed along different LoS, some LoS will show a strong blue feature while others will show none. Similarly there also is some variation for red peaks close to the line-center being suppressed given the bimodality there.

In the upcoming Section~\ref{sec:spectral_impact}, we will investigate two different quantitative measures to characterize the spectral variations for different lines of sight as implied here.

\subsection{Variations in Transmission Curves}
\label{sec:spectral_impact}
After studying the averaged transmission curves, we proceed to quantify the variations of the transmission curves. Those variations are observable features of the emerging spectra after traversing the IGM.

\subsubsection{Peak Distribution}
\label{sec:Npeaks}
As discussed in \S~\ref{sec:intro} and \S~\ref{sec:inputspectra}, observed \Lya spectra a low-$z$ exhibit usually a double or single red peaked spectrum.
Attenuation in the IGM can modify the observed peak count in some LoS.

For an
observed spectral flux density $I_\lambda(\Delta\lambda)$, we define a peak as connected
flux density bins such that $I_\lambda(\Delta \lambda)>I_T$ for a threshold $I_T$. Here we
set $I_T=0.01\cdot
\max_{0<\lambda<\infty}{\left(I_{\lambda,\mathrm{input}}\right)}$. This
criterion (while not its specific value) seeks to represent the flux sensitivity
of a generic instrument. Furthermore, we require distinct connected areas to have a
minimal separation of $0.2\ $\AA\  ($R\sim 6000$) from one another. This criterion has the purpose to not falsely
identify multiple peaks due to very small flux discontinuities that might
additionally be below the spectral resolution of the spectrograph.

\begin{figure}
\centering
  \includegraphics[width=1.0\linewidth]{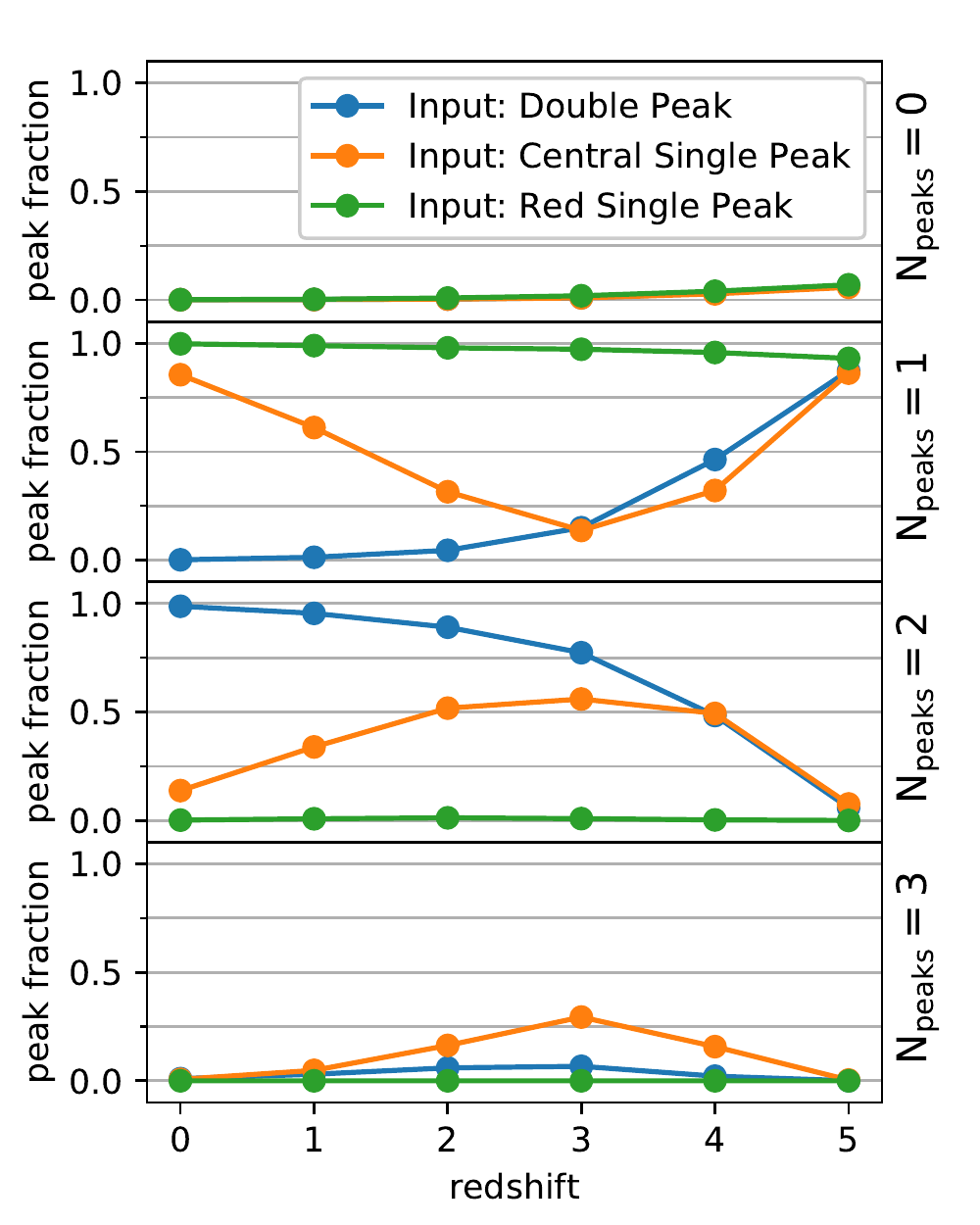}
  \caption{The evolution of detectable \Lya peaks assuming the three different intrinsic spectra as described in \S~\ref{sec:inputspectra}. The peak detection algorithm is described in \S~\ref{sec:Npeaks}.}
\label{fig:Npeaks_redshifts}
\end{figure}

Figure~\ref{fig:Npeaks_redshifts} shows the distribution of the spectral peak count $0\leq$~$\npeaks \leq 3$ given this algorithm over the redshift range from 0 to 5 for the different input spectra. Hardly any LoS exist with $\npeaks > 3$.
The evolution with redshift is anchored by the intrinsic value of \npeaks at $z\sim 0$ (i.e., $\npeaks=2$ and $1$ for the double peaked input spectrum and the other two, respectively) and a single red peak at $z=5$, while in roughly 7\% of all LoS nearly all flux and thus all peaks are suppressed at $z=5$. For the intermediate redshifts, the rugged transmission curve (cf. \S~\ref{sec:transmission}) causes a substantial increase in the number of observable peaks for the wide, central input spectrum mimicking a porous ISM. Specifically, the number of resultant double and even triple peaks increases to $\sim 50\%$ and $30\%$ at $z\sim 2-4$, respectively. Thus, the triple peaked case is
crucial in differentiating possible scenarios for the small-scale input spectra.

\subsubsection{Blue Peak Flux}
\label{sec:BluePeakFlux}
We introduce two observables to quantify the peak asymmetry. Namely the flux ratio $L_\mathrm{ratio}$ between the integrated flux $L_\mathrm{blue}$ for wavelengths below the line-center and the total observed flux $(L_\mathrm{blue}+L_\mathrm{red})$: 

\begin{align}
    L_\mathrm{ratio} \equiv \frac{L_\mathrm{blue}}{L_\mathrm{blue}+L_\mathrm{red}}= \frac{\int_{0}^{\lambda_c} T(\lambda) \cdot I(\lambda) d\lambda}{\int_{0}^{\infty} T(\lambda) \cdot I(\lambda) d\lambda}.
\end{align}

Analogously, we define the peak flux ratio $F_\mathrm{ratio}$ as the ratio of maximal flux blueward to the sum of the peak fluxes on both sides of the line-center, i.e.:

\begin{align}
    F_\mathrm{ratio} \equiv \frac{F_\mathrm{blue}}{F_\mathrm{blue}+F_\mathrm{red}} =  \frac{\underset{{0<\lambda<\lambda_c}}{\max}\left(T(\lambda) \cdot I\right)}{\underset{{0<\lambda<\lambda_c}}{\max}\left(T(\lambda) \cdot I\right)+\underset{{\lambda_c<\lambda<\infty}}{\max}\left(T(\lambda) \cdot I\right)}.
\end{align}

Note that observational studies used similar measures in the past \citep[e.g.,][]{Erb2014,Verhamme2017}.

\begin{figure}
\centering
  \includegraphics[width=1.0\linewidth]{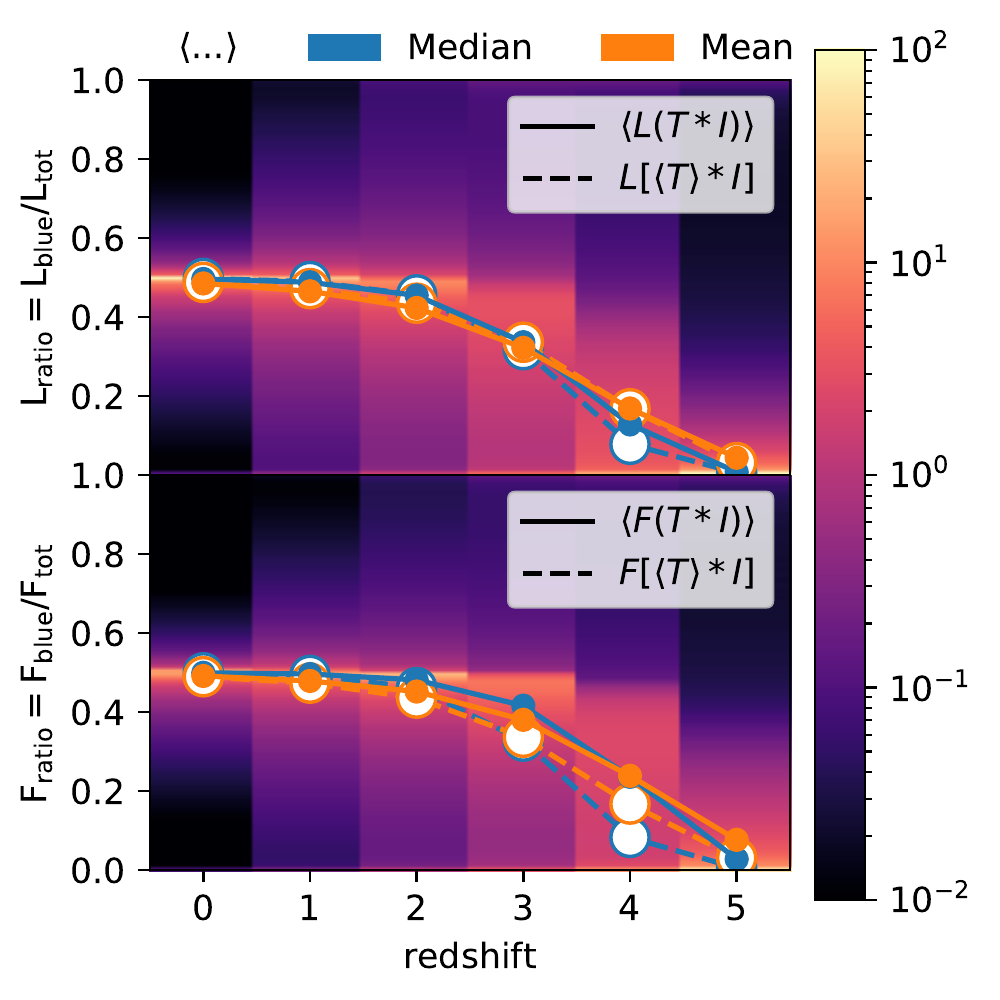}
  \caption{The ratios $L_{\mathrm{ratio}}$ (\textbf{top}) and $F_{\mathrm{ratio}}$ (\textbf{bottom}) are shown as a PDF over all lines of sight for both all directions and emitters at a given redshift as shown by the colormap. We chose the double peaked input spectrum, but find a very similar result for the centrally peaked input spectrum. The four overplotted lines in each panel show different averaging methods (see text). We assume that no peak can be detected (hence no ratio) when the flux remaining after passing the IGM is less than 1\% of the intrinsic flux $I(\lambda)$. 
  }
\label{fig:LandFstats}
\end{figure}

In Figure~\ref{fig:LandFstats}, we show the distributions of $L_{\mathrm{ratio}}$ and $F_{\mathrm{ratio}}$ across all LoS (both directions and emitters) using the double peaked intrinsic spectrum introduced in \S~\ref{sec:inputspectra}. The distribution looks very similar when using the Gaussian input spectrum.
Note that additionally we impose a minimum flux on each side of the line-center of 1\% the flux of the input spectrum for a line of sight to be deemed detectable. Given the symmetric input spectrum, a ratio of $0.5$ signifies an equal impact of the IGM on the blue and red side of the line-center. Thus, while Figure~\ref{fig:LandFstats} makes a prediction for the observed distribution given the idealized input spectrum, it also represents the IGM's impact on the asymmetry as such, with values greater than $0.5$ indicating a favorable escape of blue photons through the IGM and less than $0.5$ favoring red photons.

The solid lines show the mean and the median for the ratios, while the dashed lines show ratio for the mean and median transmission curves $\langle T(\Delta\lambda)\rangle$ multiplied by the input spectra. Both ratios intuitively follow the expected redshift evolution for all lines: At redshift $z=0$ the asymmetry is mostly unaffected by the IGM, while at higher redshifts the averaged ratios drop towards zero at $z=5$ as the IGM becomes more opaque due to a higher physical neutral hydrogen density. A closer look at the peak asymmetry distribution at a given redshift reveals a more nuanced picture: For example, at low to intermediate redshifts ($z\lesssim 3$) the distribution appears somewhat symmetric around the median with increasing variance for higher redshifts. This means that one might surprisingly find a dominant blue peak at high redshifts even though the intrinsic spectrum
is asymmetric towards the red (e.g., due to galactic outflows).
At $z\gtrsim4$ the distribution becomes positively skewed thus still allowing a range of spectra containing significant blue contributions.
This is another reason why the use of the averaged transmission curves could be deceiving about the underlying peak asymmetry distribution and thus the occurrence of such ratio. For instance for $z=5.0$ the median transmission curve leads to a $L_\mathrm{ratio}$ on sub-percent level, giving the perception that blue peaks are singularities at such redshift, when in reality we find roughly 10\% of \Lya emitting galaxies still showing significant blue flux ($L_\mathrm{blue}\geq 0.25\cdot L_\mathrm{red}$) (if present intrinsically).

\subsubsection{Large-Scale Environment}
\begin{figure}
\centering
  \includegraphics[width=1.0\linewidth]{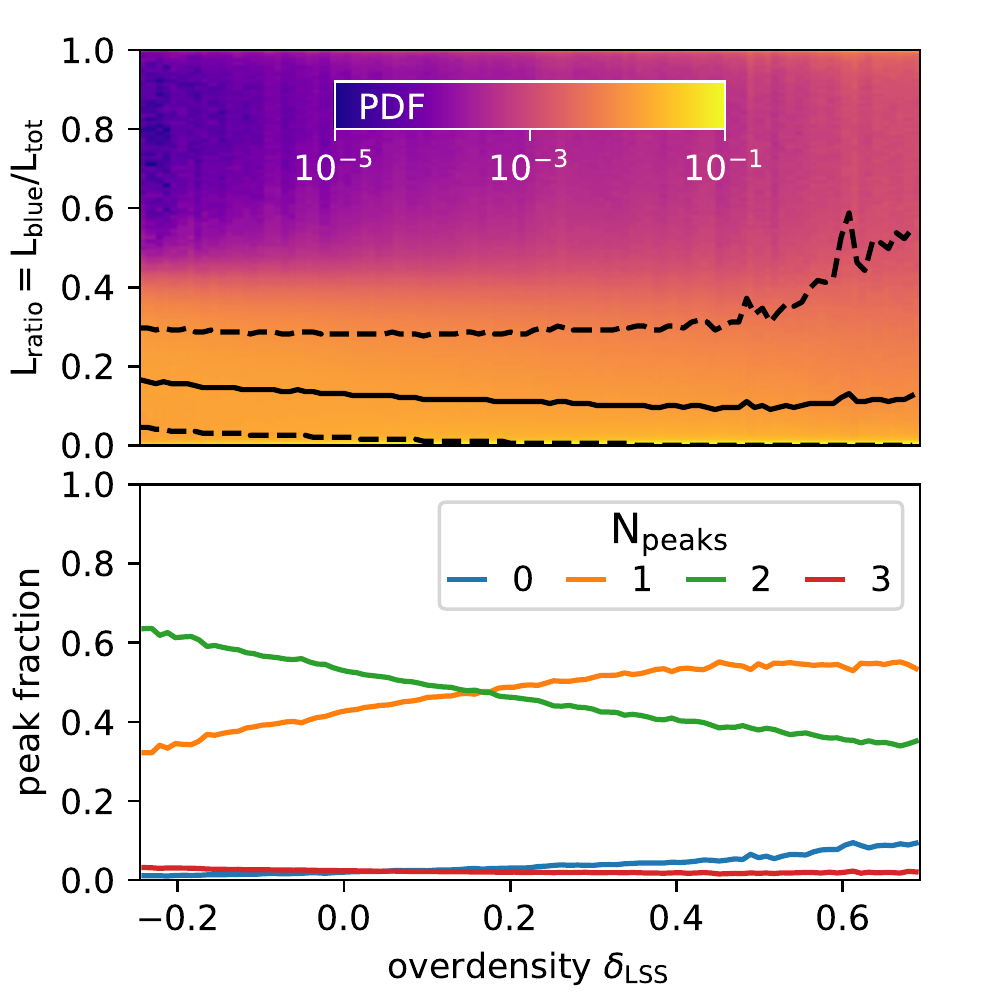}
  \caption{The $L_{\mathrm{ratio}}$ and \npeaks statistics shown as a function of the large scale overdensity $\delta_\mathrm{LSS}$ at redshift $z=4.0$. The input spectrum is the Neufeld solution. The colormap in the upper panel shows the underlying PDF normalized at a given $\delta_\mathrm{LSS}$. Overdense regions overall decrease $L_{\mathrm{ratio}}$. At higher ovedensity this trend halts as the ratio becomes more fluctuating for different LoS. For the peak fractions we find a gradual decrease of double peaks. This is caused by absorption features on the blue side of the line-center. Thus the decrease of double peaks is strongly correlated with the increase of the fraction of single peaked spectra.}
\label{fig:LandFstats_LSScorr_z4_overdensity}
\end{figure}

Beside redshift and input spectrum, we find that our proposed statistics also depend on the large-scale environment that the emitters reside in. Most prominently we find a correlation of the flux ratio and peak fraction with the linear overdensity as shown in Figure~\ref{fig:LandFstats_LSScorr_z4_overdensity} for $z=4.0$. We calculate the overdensity using a Gaussian smoothing kernel with $\sigma=3\ $cMpc/h. The flux ratio slightly decreases towards higher overdensities indicating more intervening neutral hydrogen. At the same time, the scatter of the flux ratio strongly increases as more varying matter structure is passed along the lines of sight. The fraction of double peaks, as present in the input spectrum, strongly decreases towards higher overdensities as more and more blue peaks are attenuated. In up to 10\% of the cases, the red peak is additionally suppressed, completely obscuring those \Lya emitting galaxies in high overdensity environments. This also affects the clustering signal of \Lya emitters as has been studied in~\citet{Zheng2011} and~\citet{Behrens2018} in more detail.  
The correlation with overdensity is much weaker for redshift between $z=0.0$ and $2.0$ where the most likely outcome remains $L_{\mathrm{ratio}}=0.5$ over the overdensity range. This behaviour changes at $z=3.0$ in overdense regions, where the distribution starts to get skewed towards $L_{\mathrm{ratio}}$ close to zero. There is a similar overall correlation at $z=3.0$ and $z=4.0$. At higher redshifts, the correlation appears to be smaller as most blue peaks are already attenuated even in underdense regions.

\section{Conclusions}
The \Lya line can be used to constrain the neutral hydrogen distribution from galactic to cosmological scales. This versatility is, however, also a curse since degeneracies between these scales exist
that come into play at $z\gtrsim 3$ when the more opaque IGM can compensate the effects of a more porous ISM.
Such a scenario -- which is suggested by cosmological simulations -- allows not only the escape of \Lya photons closer to line-center but also of ionizing photons
that are susceptible to the same galactic HI distribution
\citep[e.g.][]{Dijkstra2016}.

Using a large set of transmission curves we calculate from the IllustrisTNG100 simulations, we have quantified the impact of the IGM on spectra for $z=0-5$. 
Doing so, we study a new approach to break the degeneracy using two statistics, namely the peak count and peak asymmetry (\S~\ref{sec:Npeaks}/~\ref{sec:BluePeakFlux}). 
In particular, we found the fraction of triple peaks to be an important differentiator 
for different intrinsic spectra. 
In contrast, we show that the commonly used averaged transmission curves can be deceiving for the interpretation of the IGM's impact on observed \Lya spectra and their redshift evolution. 

Making our catalogs of transmission curves publicly available \citep{Byrohl2020dataset} allows others to incorporate a better IGM treatment and its impact on \Lya spectra. At the same time, this flexibly allows to refine presented statistics and the intrinsic spectral modeling in the future.

Our findings require a high spectral resolution (optimally $R\gtrsim 6000$), particularly for the peak count statistic. 
Hence, current samples of \Lya spectra at $z\gtrsim 3$ possess a too low spectral resolution in order to test our predictions \citep[e.g.,][]{Erb2014,Herenz2017}. However, individual triple-peaked spectra have been observed \citep[e.g.][]{2017A&A...608L...4R,Vanzella2019}, and
future surveys will be able to increase this count to a statistical sample that allows to break the above described degeneracy.

\begin{acknowledgements}
We thank Christoph Behrens for providing us with the early version of \textit{ILTIS},
Eiichiro Komatsu and Shun Saito for useful discussions on this draft, and 
Dylan Nelson for his help processing the IllustrisTNG data. 
The radiative transfer simulations and analysis were conducted on the supercomputers at the Max Planck Max Planck Computing and Data Facility (MPCDF). We acknowledge use of the Python programming language \citep{VanRossum1991}, and use of the  Astropy \citep{astropy}, Numpy \citep{VanDerWalt2011}, IPython \citep{Perez2007}, Dask \citep{dask}, h5py \citep{h5py} and Matplotlib \citep{Hunter2007} packages for post-processing.
       MG was supported by NASA through the NASA Hubble Fellowship grant HST-HF2-51409 and acknowledges support from HST grants HST-GO-15643.017-A, HST-AR-15039.003-A, and XSEDE grant TG-AST180036.
 
\end{acknowledgements}

\bibliographystyle{mnras}
\bibliography{references}

\appendix
\section{Spectral Resolution}
\label{sec:spectralres}

\begin{figure}[htb]
\centering
  \includegraphics[width=1.0\linewidth]{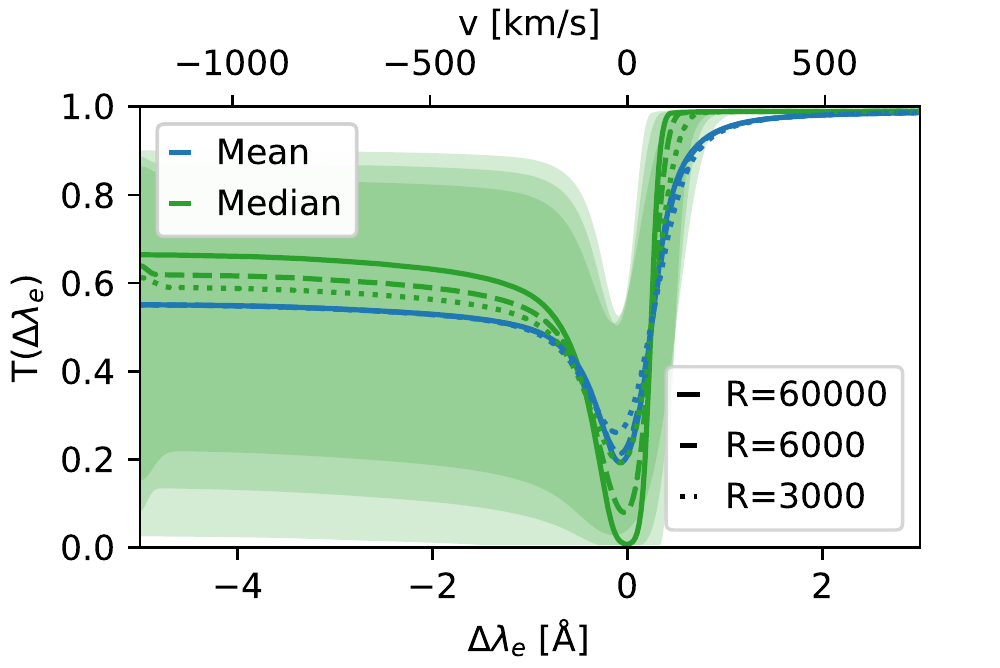}
  \caption{Impact of the spectral resolution on the PDFs of the transmission function $T$ as a function of input wavelength shift $\Delta\lambda_e$. We show the PDF of $T$ over all emitters and LoS at $z=3$ similar to Figure~\ref{fig:transmissionPDFz3}. In blue we show the mean for a given $\Delta\lambda_e$, while in green we show the median. We show the transmission at different spectral resolutions $R\in\left\{60000,6000,3000\right\}$ (solid,dashed,dotted). The shaded regions enclose the central $68\%$ of all individual transmission curves. The darkest (lightest) shade corresponds to the lowest (highest) spectral resolution.}
\label{fig:transmission_z3_spectralres}
\end{figure}

In Figure~\ref{fig:transmission_z3_spectralres} we show the impact of the spectral resolution on the averaged transmission curve at $z=3$. We show the median along with the 16th and 84th percentile and the mean over the different emitters and LoS for at a given wavelength offsets $\Delta\lambda_e$. We find that the mean is nearly independent of the spectral resolution, while the median and percentiles in general strongly dependent on chosen resolution. This is easily explained:
Linearity, i.e. $\mathrm{STAT}\left[A+B\right]=\mathrm{STAT}\left[A\right]+\mathrm{STAT}\left[B\right]$ for two distributions $A$ and $B$ and a summary statistic $\mathrm{STAT}$, holds for the mean but not for percentiles such as the median. Take, for example, two PDFs $A=P(T)_{\Delta\lambda_e=\lambda_{i}}$ and $B=P(T)_{\Delta\lambda_e=\lambda_j}$, where $\lambda_{i}$ and $\lambda_j$ describe two neighboring wavelength bins. The mean transmissivity T over those two bins can be determined from the mean of $A$ and $B$ alone, while this is not possible for the median. In particular if two neighboring bins have the same mean, the mean of the sum of those distributions has to be the same, while this is not true for the median as can be readily seen in Figure~\ref{fig:transmission_z3_spectralres}. This behaviour is furthermore complicated for the median as the distributions in neighbouring bins are strongly correlated. As a result of this, we recommend using the mean transmission curves over the median unless the spectral resolution is clearly stated.

\label{lastpage}
\end{document}